\documentclass[aps,prc,twocolumn,showpacs,superscriptaddress]{revtex4-1}

\usepackage{subfigure}
\usepackage{graphicx, color}
\usepackage{float}
\usepackage{hyperref}
\usepackage[all]{hypcap}
\usepackage{array}
\usepackage{graphicx}
\usepackage{amsmath,amsthm,amssymb}
\usepackage{color}

\begin{document}
\title{Medium-induced flavor conversion and kaon spectra in electron-ion collisions}

\author{Ning-Bo Chang}
\affiliation{Key Laboratory of Quark and Lepton Physics (MOE) and Institute of Particle Physics, Central China Normal University, Wuhan 430079, China}

\author{Wei-Tian Deng}
\affiliation{School of Physics, Huazhong University of Science and Technology, Wuhan 430074, China}

\author{Xin-Nian Wang}
\affiliation{Key Laboratory of Quark and Lepton Physics (MOE) and Institute of Particle Physics, Central China Normal University, Wuhan 430079, China}
\affiliation{Nuclear Science Division Mailstop 70R0319,  Lawrence Berkeley National Laboratory, Berkeley, California 94740, USA}

\date{\today}
\pacs{24.85.+p, 12.38.Bx, 13.87.Ce, 13.60.-r}

\begin{abstract}
Multiple scattering and induced parton splitting lead to a medium modification of the  QCD evolution for jet fragmentation functions and the final hadron spectra. Medium-induced parton splittings not only lead to energy loss of leading partons and suppression of leading hadron spectra, but also modify the flavor composition of a jet due to induced flavor conversion via gluon emission, quark pair production and annihilation.  Through a numerical study of the medium-modified QCD evolution, leading $K^-$ strange meson spectra are found to be particularly sensitive to the induced flavor conversion in semi-inclusive deeply inelastic scatterings (SIDIS) off a large nucleus. The induced flavor conversion can lead to increased number of gluons and sea quarks in a jet and, as a consequence, enhance the leading $K^-$ spectra to counter the effect of parton energy loss in SIDIS with large momentum fractions $x_B$ where the struck quarks are mostly valence quarks of the nucleus.
\end{abstract}

\maketitle

\section{Introduction}
 
In the study of quark gluon plasma (QGP) produced in relativistic heavy-ion collisions, jet quenching has been established as one of the evidences for the formation of strongly coupled QGP in experiments at both the Relativistic Heavy-ion Collider (RHIC) and  the Large Hadron Collider (LHC) \cite{Jacobs:2004qv,Majumder:2010qh}. The phenomenon as manifested in the suppression of jets and final hadron spectra at large transverse momentum is a consequence of parton energy loss induced by multiple scatterings during the propagation of the initial hard partons through the dense medium of QGP \cite{Bjorken:1982tu,Gyulassy:1993hr,Baier:1994bd,Zakharov:1996fv,Gyulassy:2000fs,Wiedemann:2000za,Arnold:2001ba,Guo:2000nz}. A similar phenomenon is also predicted to happen in semi-inclusive deeply inelastic scatterings (SIDIS) off a large nucleus \cite{Guo:2000nz,Wang:2002ri,Arleo:2003jz} in which multiple scattering between the struck quark and the cold nuclear medium can lead to suppression of the final leading hadron spectra \cite{Ashman:1991cx,Hafidi:2006ig,Airapetian:2007vu,Airapetian:2012ki}.  Phenomenological studies of experimental data on parton energy loss can yield important information on the properties of the hot or cold nuclear medium \cite{Burke:2013yra}.

The suppression of final large transverse momentum hadron spectra in heavy-ion collisions and leading hadron spectra in SIDIS can both be described by a medium modification of the fragmentation functions (FF's) of a propagating parton. The modification is governed by a set of medium-modified QCD evolution equations that incorporate multiple parton scattering and induced gluon bremsstrahlung through medium-modified parton splitting functions \cite{Wang:2002ri,Wang:2009qb,Deng:2010xv,Chang:2014fba}. These medium-modified splitting functions lead to energy loss of the leading parton within a jet and suppression of hadrons with large fractional momenta in the final jet. They also give rise to an enhancement of soft or non-leading hadrons through hadronization of the radiated gluons and partons from induced pair production. Together with the flavor conversion processes of leading partons \cite{Guo:2000nz,Liu:2006sf,Liu:2008zb}, induced radiation and pair production can also change the hadron flavor composition of the modified jets. Inclusion of these flavor conversion processes is critical for a precise description of jet quenching effects on identified hadron spectra and extraction of medium  properties in high-energy heavy-ion and electron-ion collisions. 

In this article, we will study the effect of induced flavor conversion on final leading hadron spectra in SIDIS off a large nucleus within the framework of the high-twist approach to multiple parton scatterings and modified parton fragmentation functions. We will numerically solve a set of medium-modified Dokshitzer-Gribov-Lipatov-Altarelli-Parisi (mDGLAP) evolution equations for parton fragmentation functions that include flavor conversion through induced gluon emission and pair production. We examine in detail how identified hadron spectra are modified. For moderate and large values of the struck quark momentum fraction $x_B$, the fragmenting partons in SIDIS are dominated by valence quarks ($u$ and $d$) from the target nucleus. Since constituent quarks in $K^-$ strange mesons ($s$, $\bar u$) can only come from pair production in the shower evolution of the $u$ and $d$ quark, their spectra are particularly sensitive to the medium-induced flavor conversion during the propagation of the struck valence quark in the nucleus. We will show that the medium-induced flavor conversion can enhance $K^-$ spectra with large momentum fraction to counter the effect of parton energy loss in SIDIS with moderate and large values of $x_B$. 

This article is organized as follows.  In Sect.~\ref{sec:ini}, we briefly review our framework for medium modified fragmentation functions (mFF's) in terms of medium-modified QCD evolution equations. In Sect.~\ref{sec:dis}, we present numerical solutions for the medium modified fragmentation functions and in particular the enhancement of $K^-$ spectra in some  selected kinematic regions. We will also discuss theoretical errors from parameterizations of the vacuum FF's and nuclear parton distribution functions (nPDF). Finally a summary and discussions are given in Sect.~\ref{sec:sum}.

\section{Medium modified fragmentation functions}
\label{sec:ini}

Within generalized collinear factorization \cite{Qiu:1990xy,Kang:2013raa,Kang:2014ela}, 
one can calculate contributions to the differential cross section of 
SIDIS, $e^- (L)+ A(p) \rightarrow e^-(L') + h(p_h)+X$, from the leading order (LO) high-twist processes in which 
a quark from the target nucleus is struck by the virtual photon $\gamma^*(q)$ with momentum $q=L-L'=(-Q^2/2q^-, q^- , \vec 0_\perp)$ and virtuality $Q^2$. In the infinite momentum frame, each nucleon inside the target nucleus carries a momentum $p=(p^+,0,\vec 0_\perp)$. The struck quark with an initial momentum fraction $x_B=Q^2/2p^+q^-$ then undergoes  multiple scatterings with the remnant of the  target nucleus before fragmenting into a final hadron with momentum $p_h=(0,z_hq^-,\vec 0_\perp)$. The final differential cross section at LO can be expressed in terms of a collinear hard part of photon-quark scattering and a medium-modified fragmentation function (mFF) $\widetilde D_q^h(z_h,Q^2)$ \cite{Guo:2000nz}. 

Following the approach for evolution of the renormalized FF's in vacuum, we can also sum the leading-log and twist-four medium corrections and arrive at the mDGLAP evolution equations for mFF's \cite{Guo:2000nz,Wang:2009qb,Schafer:2007xh},
\begin{eqnarray}
 \label{mDGLAP1}
 \frac{\partial \tilde{D}_q^h(z_h,Q^2)}{\partial \ln Q^2}\hspace{-4pt} &=&\hspace{-4pt}\frac{\alpha_s(Q^2)}{2\pi}\hspace{-4pt} \int_{z_h}^1
 \frac{dz}{z}\left [ \tilde{\gamma}_{q\rightarrow qg}(z,Q^2)\tilde{D}_q^h(\frac{z_h}{z},Q^2)\right. \nonumber \\
 &+&\left.\tilde{\gamma}_{q\rightarrow gq}(z,Q^2)\tilde{D}_g^h(\frac{z_h}{z},Q^2)\right ] ,\\
\label{mDGLAP2}
  \frac{\partial \tilde{D}_g^h(z_h,Q^2)}{\partial\ln Q^2}\hspace{-4pt}&=&\hspace{-4pt}\frac{\alpha_s(Q^2)}{2\pi} \hspace{-4pt} \int_{z_h}^1
 \frac{dz}{z}\left [ \tilde{\gamma}_{g\rightarrow gg}(z,Q^2)\tilde{D}_g^h(\frac{z_h}{z},Q^2) \right.  \nonumber \\
 &+&\left. \sum_{q=1}^{2n_f}\tilde{\gamma}_{g\rightarrow q\bar q}(z,Q^2)\tilde{D}_q^h(\frac{z_h}{z},Q^2)\right ] ,
\end{eqnarray}
which are similar to the DGLAP equations for vacuum FF's. The differences here from the vacuum DGLAP equations are the medium-modified splitting functions $\tilde \gamma _{a\rightarrow bc}$,
\begin{equation}
 \tilde \gamma _{a\rightarrow bc}(z,Q^2)=\gamma_{a\rightarrow bc}(z)+ \Delta \gamma_{a\rightarrow bc}(z,Q^2),
\end{equation}
which contain both the parton splittings in vacuum $\gamma_{a\rightarrow bc}(z)$ and medium-induced ones $\Delta \gamma_{a\rightarrow bc}(z,Q^2)$, whose detailed expressions can be found in Refs.~\cite{Wang:2009qb,Schafer:2007xh}.  Note that the medium-induced splitting functions  $\Delta \gamma_{a\rightarrow bc}$ depend on the jet transport parameter $\hat q$ integrated over the path length of the quark propagation. For example,
\begin{eqnarray}
 \Delta \gamma_{q\rightarrow qg}(z,\ell_T^2)&=&\frac{1}{\ell_T^2+\mu_D^2}\left[ C_{A}(1-z)(1+(1-z)^{2})/z \right. \nonumber \\
 &+& \left. C_{F}z(1+(1-z)^{2})\right] \nonumber \\
 &\times&  \int dy^-\hat q(y^-) 4 \sin^{2}(x_{L}p^{+}y^{-}/2),
 \label{delta_sp_qqg} 
\end{eqnarray}
is the medium-induced quark splitting function, where $\ell_T$ is the relative transverse momentum of the final partons and $x_L=\ell_T^2/2p^+q^-z(1-z)$ is the fractional light-cone momentum carried by the hard parton from the target nucleus that induces the parton splitting, $y^-$ is the light-cone coordinate of the target nucleons involved in the secondary scattering and $\mu_D$ represents beam partons' average intrinsic transverse momentum inside a nucleon. The jet transport parameter $\hat q$ here arises from the twist-four quark-gluon correlation function in a factorized form as assumed in Ref.~\cite{Wang:2009qb}.  In the next-to-leading order (NLO) calculation of transverse momentum broadening in SIDIS \cite{Kang:2013raa,Kang:2014ela} , one can derive the QCD evolution equation for $\hat q$. The energy and scale dependence of $\hat q$ from such evolution equation can be used together with the NLO results to provide a consistent prediction of the energy and scale dependence of the transverse momentum broadening of final hadrons in SIDIS. Within the LO high-twist approach, we will assume, however, a constant $\hat q$ for the phenomenological study in this article.
 The medium-induced quark-to-gluon and gluon-to-quark splitting functions in Eqs.~(\ref{mDGLAP1}) and (\ref{mDGLAP2})  couple the quark and gluon fragmentation functions through the mDGLAP equations. These are where the medium-induced flavor conversion occurs and will lead to a change in the flavor composition in the mFF's of the quark jets in SIDIS. 

To solve the mDGLAP evolution equations  in Eqs.~(\ref{mDGLAP1}) and (\ref{mDGLAP2}), we have to provide the initial conditions of mFF's at a given initial scale $Q_0$. These initial conditions are not calculable in perturbative QCD (pQCD) and have to be given by a model assumption. Instead of using vacuum FF's for the initial conditions \cite{Majumder:2013re}, we proposed a {\it convoluted} model \cite{Chang:2014fba} in order to take into account of parton energy loss for partons with virtualities below $Q_0^2$. The {\it convoluted} initial conditions are obtained from the convolution of vacuum FF's $D_a^h(z,Q^2)$ at the initial scale $Q_0^2$ and the quenching weight due to induced gluon radiation,
 \begin{eqnarray}
  \widetilde D_g^h(z,Q_0^2)&=&\int_0^1 d\epsilon P_g(\epsilon,Q_0^2)
  \frac{1}{1-\epsilon} D_g^h(\frac{z}{1-\epsilon},Q_0^2) \nonumber \\
  &+& \int_0^1 d\epsilon G^g(\epsilon,Q_0^2)\frac{1}{\epsilon}D^h_g(\frac{z}{\epsilon},Q_0^2)\,,\hspace{0.3in}   \label{eqquark}\\
  \widetilde D_q^h(z,Q_0^2)&=&\int_0^1 d\epsilon P_q(\epsilon,Q_0^2)\frac{1}{1-\epsilon} D_q^h(\frac{z}{1-\epsilon},Q_0^2) \nonumber \\
  &+& \int_0^1 d\epsilon G^q(\epsilon,Q_0^2)\frac{1}{\epsilon} D_g^h(\frac{z}{\epsilon},Q_0^2),\hspace{0.3in} 
   \label{eqgluon}
\end{eqnarray} 
where the quenching weight $P_a(\epsilon,Q_0^2)$ represents the probability of total fractional energy loss $\epsilon$ by the initial parton $a$ due to induced gluon radiation and $G^a(\epsilon)$ represents the spectrum distribution of the radiated gluons with fractional energy $\epsilon$. 
The quenching weight  $P_a(\epsilon,Q_0^2)$ is calculated from a Poisson convolution of the single gluon spectrum $dN_g^a/dz$ at scale $Q_0^2$,
\begin{eqnarray}
  P_a(\epsilon,Q_0^2)&=&\sum_{n=0}^\infty \frac{1}{n!}
 \prod_{i=1}^n \int_0^{1} dz_i \frac{dN_g^a}{dz_i}(Q_0^2)
    \delta(\epsilon-\sum_{i=1}^n z_i)\nonumber \\
    & & \times \exp\left[ - \int_0^1 dz\frac{dN_g^a}{dz}(Q_0^2)\right]\, ,
   \label{eq:pdeltaeps}
\end{eqnarray}
under the assumption that the number of independent gluon emissions satisfies the Poisson distribution. We use Monte Carlo simulations to calculate the quenching weight $P_a(\epsilon,Q_0^2)$. This method also enables us to record the energy fraction of each radiated gluon and then obtain the gluon energy spectrum $G^a(\epsilon)$ from multiple induced emissions. With $G^a(\epsilon)$,  we can include contributions from the fragmentation of radiated gluons to the initial conditions and also ensure the momentum conservation at the same time. Using such initial conditions for the mDGLAP equations, we can describe the HERMES data \cite{Airapetian:2007vu} better as compared to other models for initial conditions. Details can be found in Ref.~\cite{Chang:2014fba}.

\section{Medium-induced flavor conversion in SIDIS}
\label{sec:dis}

Jet quenching in SIDIS is measured experimentally via the suppression of leading hadron spectra. The nuclear modification factor $R_A^h$ for hadron spectra is defined in terms of a ratio of hadron yields per DIS event $N^h/N^e$ for a nuclear target $A$ to that for a deuterium target $D$ \cite{Ashman:1991cx,Hafidi:2006ig,Airapetian:2007vu,Airapetian:2012ki},
\begin{eqnarray}
R_A^h(\nu,Q^2,z)=\left[\frac{N^h(\nu,Q^2,z)}{N^e(\nu,Q^2)}\right]_A / \left[\frac{N^h(\nu,Q^2,z)}{N^e(\nu,Q^2)}\right]_D.
\label{eq:R_a}
\end{eqnarray}
Hadron yields per DIS event $N^h/N^e$ from LO pQCD can be related to the nuclear modified FF's $\widetilde{D}_q^h(z,Q^2)$ from the mDGLAP evolution equations in Eqs.~(\ref{mDGLAP1}) and (\ref{mDGLAP2}),
\begin{eqnarray}
\left. \frac{N^h(\nu,Q^2,z)}{N^e(\nu,Q^2)}\right|_A =\left. \frac{\Sigma e_q^2q(x_B,Q^2)\widetilde D_q^h(z,Q^2)}{\Sigma e_q^2q(x_B,Q^2)}\right|_A,
\label{yield-ratio}
\end{eqnarray}
where $q(x,Q^2)$ is the quark distribution function inside the nucleus and $e_q$ is the quark's charge. The mFF's $\widetilde D_q^h(z,Q^2)$ are obtained from the numerical solution of the mDGLAP equations for a given propagation path or interaction point of the photon-quark scattering inside the nucleus which is then averaged over the nucleus' volume,
\begin{equation}
 \widetilde{D}_q^h(z,Q^2)=\frac{1}{A}\int d^2b dy_0 \widetilde{D}_q^h(z,y_0,b,Q^2) \rho_A(y_0,b),
\label{eq:average_dis}
\end{equation}
where $\rho_A(y,b)$ is the Woods-Saxon nuclear density distribution which is normalized as $\int dy d^{2}b\rho_{A}(y,b)=A$. The jet transport parameter $\hat{q}$ along the propagation path of the struck quark is assumed to be proportional to the nuclear density, $\hat q(y,b)=\hat q_{0} \rho_{A}(y,b)/\rho_{A}(0,0)$, where $\hat q_{0}$ is the value of $\hat q$ at the center of the nucleus. With the {\it convoluted} initial conditions, we were able to describe the HERMES data on the hadron suppression \cite{Airapetian:2007vu} well and obtained $ \hat{q}_0= 0.020 \pm 0.005$ GeV$^2$/fm \cite{Wang:2009qb,Deng:2010xv,Chang:2014fba} from a $\chi^2$ fit.  We will use the central value of $\hat{q}_0$ from this fit for all numerical calculations in this study.

\begin{figure}[htbp]
     \includegraphics[width=3.0in]{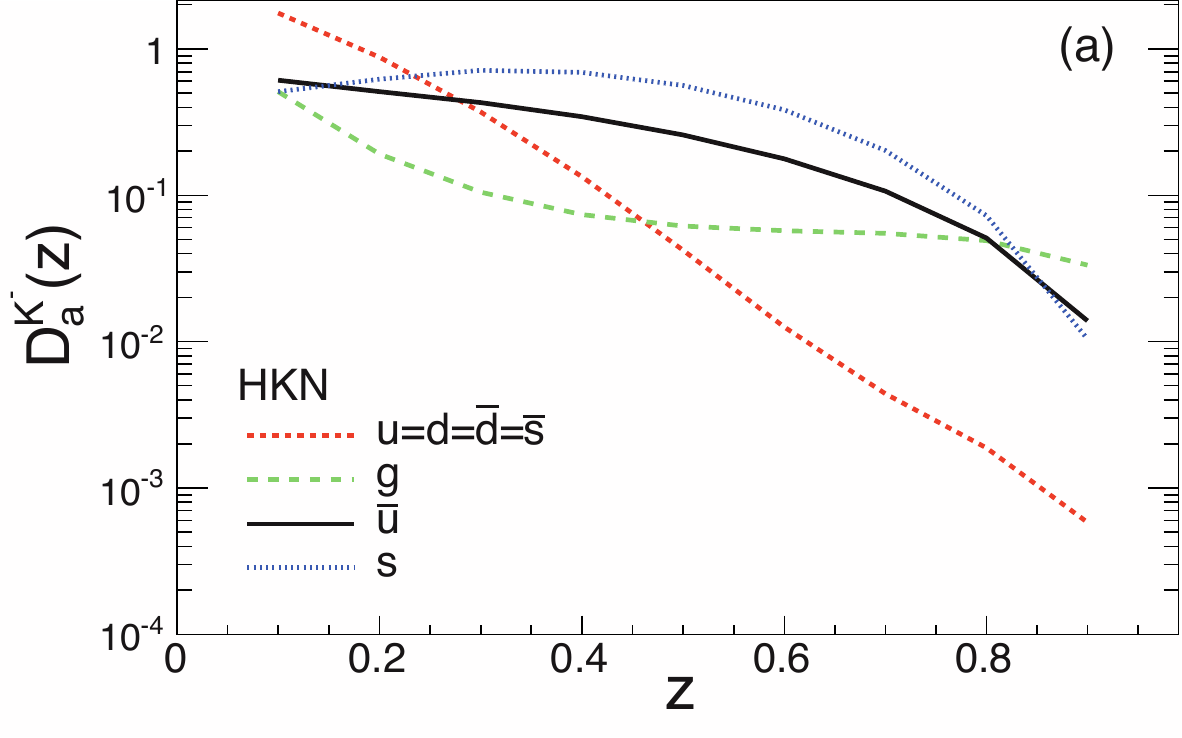}
     \includegraphics[width=3.0in]{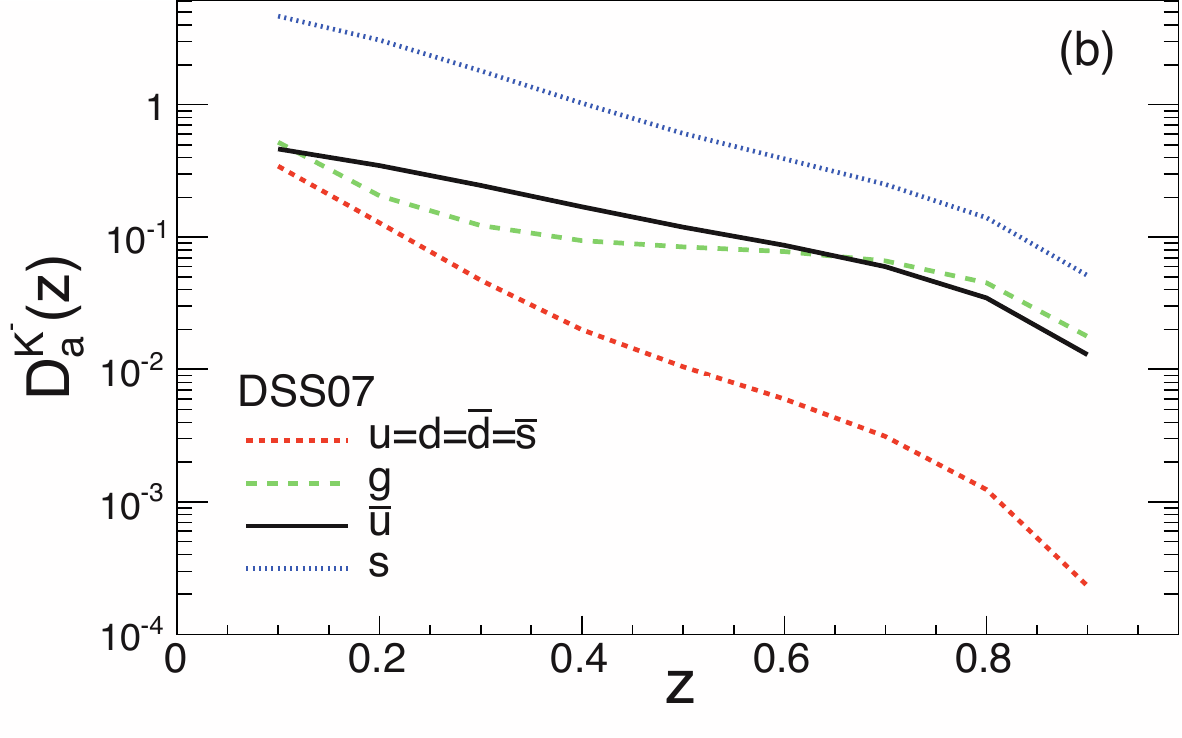}
   \caption{(color online) Parton fragmentation functions for $K^-$ in vacuum at $Q^2\approx 10$ GeV$^2$ from (a) the HKN parameterization \cite{HKN}
   and (b) DSS07 parameterization \cite{deFlorian:2007aj}.}
  \label{fig1}
\end{figure}

In search for a clear signal of the medium-induced flavor conversion in the final hadron spectra, we choose the charged strange meson $K^-$ since its constituent quarks ($s$ and $\bar u$) are not the same as any of the struck valence quarks ($u$ and $d$) of the target nucleus. In Fig.~\ref{fig1}, we show the vacuum FF's of different partons to $K^-$ as a function of the momentum fraction $z$ from two different parameterizations. The first parameterization is by Hirai, Kumano and Nagai (HKN) \cite{HKN} which was used in the past for the study of final hadron suppression due to parton energy loss in cold nuclei \cite{Wang:2009qb,Deng:2010xv,Chang:2014fba}. However, HKN parameterization of FF's is known to describe poorly the proton and kaon spectra in SIDIS off a proton target \cite{Airapetian:2012ki} mainly because of the approximate SU(3) flavor symmetry imposed on the parameterization of kaon's FF's from the constituent quarks. As seen in Fig.~\ref{fig1}(b), the deFlorian-Sassot-Stratmann (DSS07) parametrization \cite{deFlorian:2007aj} do not have such flavor symmetry and can fit the $K^+$ spectra much better \cite{Airapetian:2012ki}.  Neither of these two parameterizations can fit the $K^-$ spectra from HERMES SIDIS data perfectly, with HKN overpredicting while DSS07 underpredicting the HERMES data at intermediate and small $z$. We will use both parameterizations for our study here and consider the difference of the results as an overall theoretical uncertainty in addition to the errors in each parameterization.

\begin{figure}[htbp]
     \includegraphics[width=3.0in]{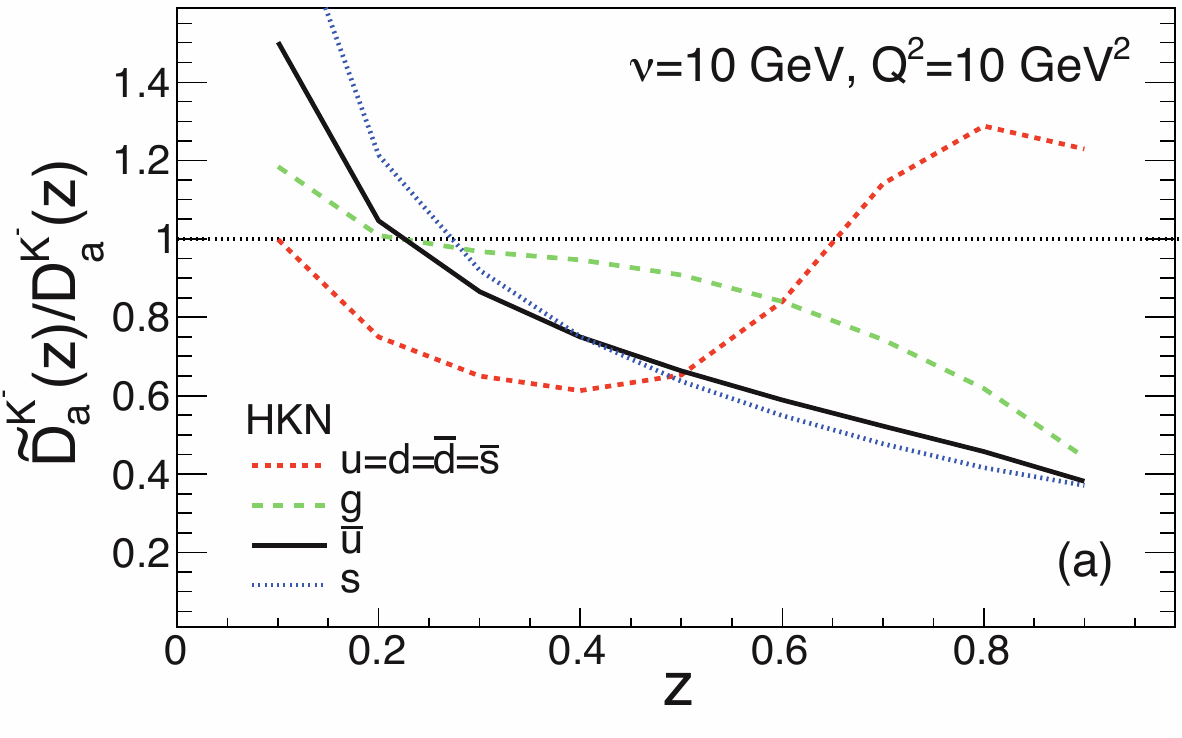}
      \includegraphics[width=3.0in]{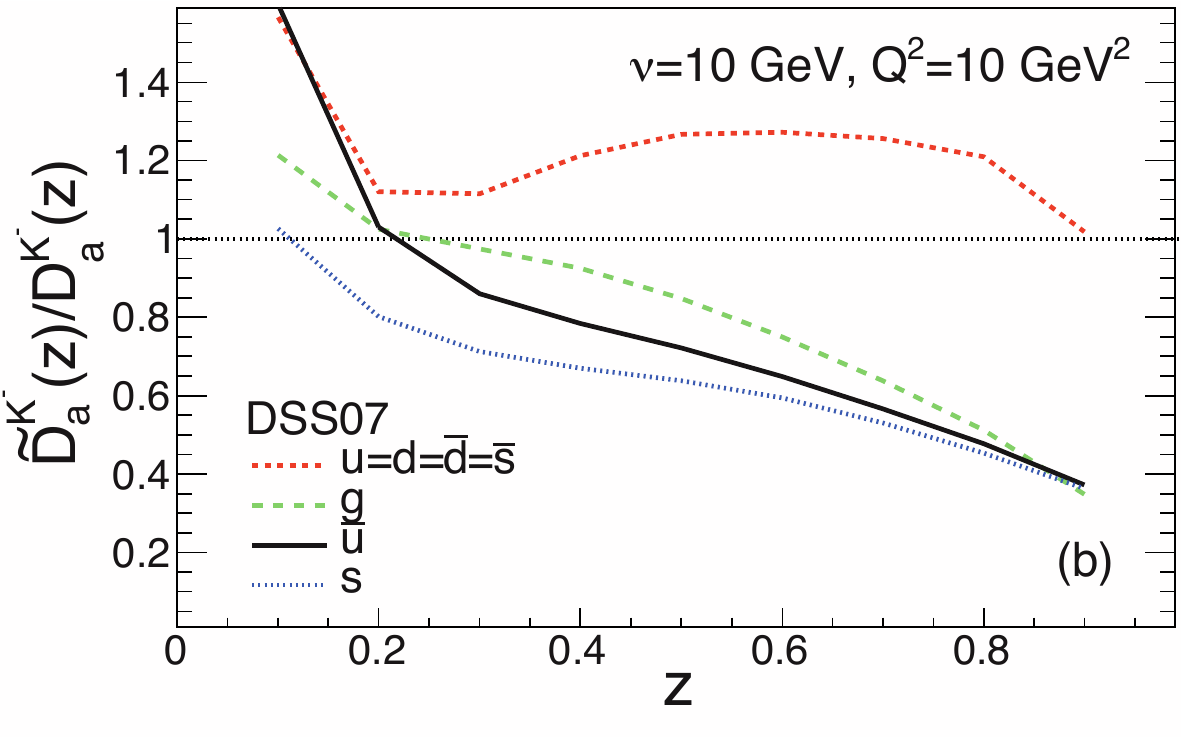}
   \caption{(color online) Ratios of mFF's to the vacuum ones for $K^-$ from different partons with initial energy $\nu=10$ GeV and $Q^2=10$ GeV$^2$ in SIDIS off a Pb target. (a) HKN and (b) DSS07 parameterization of vacuum FF's are used in these calculations.}
  \label{fig2}
\end{figure}

As seen from Fig.~\ref{fig1}, vacuum FF's of non-constituent quarks fall off exponentially at large $z$ and are much smaller than that of gluon and constituent quarks which are relatively flat.  Because of the exponential fall-off of its vacuum FF, $K^-$ production from an initial non-constituent quark should be significantly suppressed at large $z$ due to parton energy loss. On the other hand, contributions to the final $K^-$ spectrum from gluons and constituent quarks that are produced via medium-induced flavor conversion become significant. They offset the suppression of $K^-$ spectrum due to energy loss of an initial non-constituent quark and can even lead to an enhancement of the final effective mFF for $K^-$ at large $z$ as shown in Fig.~\ref{fig2}, where we plot ratios of mFF's to FF's in vacuum for $K^-$ from different initial partons.  The enhancement of $K^-$ spectra due to medium-induced flavor conversion is present in mFF's  of initial non-constituent quarks for both HKN and DSS07 parameterization of vacuum FF's. However, the enhancement is quantitatively different for these two parameterizations because of the difference in vacuum FF's for the constituent quarks ($\bar u$ and $s$). For HKN parameterization, the enhancement only appears at large $z$ and mFF's at intermediate $z$ are suppressed due to the dominance of the effect of parton energy loss of initial non-constituent quarks. For DSS07 parameterization, the enhancement is present at both large and intermediate $z$ because of the much large FF for strange quarks ($s$) and  smaller slope of the exponential fall-off of vacuum FF's at intermediate $z$ for non-constituent quarks (see Fig.~\ref{fig1}) which reduces the effect of parton energy loss.

 Parton energy loss also leads to suppression of leading $K^-$ production from an initial gluon or constituent quark jet. However, their contributions to the $K^-$ spectrum are still much larger than that from medium-induced gluon and quark pairs due to the flat behavior of their vacuum FF's. The final effective mFF's for $K^-$ from an initial gluon or constituent quark jet are therefore suppressed at both large and intermediate values of $z$ as seen in Fig.~\ref{fig2} for both parameterizations of vacuum FF's. At small $z$, all mFF's are enhanced by the medium-induced gluon emission and quark-pair production.
 
To investigate the sensitivity of charged kaon spectra to the medium-induced flavor conversion in SIDIS off a nucleus, we focus on SIDIS off a Pb nucleus at moderate and large $x_B$ so that the struck quarks are mostly valence quarks ($u$ and $d$) from the nucleus. In this region of $x_B$, the quark distribution in a nucleon is indeed dominated by valence quarks according to the CTEQ6 parameterization \cite{Pumplin:2002vw} as shown in Table~\ref{tab1}.  We should point out that the strange quark distribution at large values of $x_B$ from CTEQ6 parameterization is shown to be significantly larger than the most recent extraction from HERMES experimental data on multiplicities of strange kaons in SIDIS \cite{Airapetian:2013zaw}. Even though such change of strange quark distribution will not significantly influence our results on the nuclear modification of $K^-$ spectra in this study, one should use future parameterizations that take this into account once it is available. We also for the moment neglect the effect of nuclear modification of the parton distributions functions on the hadron yields per DIS event and simply use the vacuum parton distribution functions.

\begin{table}[h]
\begin{tabular}{|c|c|c|c|c|c|c|c|c|c|c|}
\hline
$q(x_B,Q^2)$ &$\bar{s}, s $ & $\bar{d}$ &$\bar{u}$ &u  &d  \\
\hline
$x_B=0.5$&0.0018& 0.0029& 0.0053& 0.5331& 0.1532  \\
\hline
$x_B=0.1$& 0.4790& 1.3961& 0.9262 &  5.6736& 3.7867\\
\hline
\end{tabular}\caption{Values of quark distributions inside a proton from the CTEQ6 parameterization \cite{Pumplin:2002vw} at $Q^2$=10 and 2 GeV$^2$ for $x_B=0.5$ and $x_B=0.1$, respectively.}
\label{tab1} 
\end{table}

\begin{center}
\begin{figure}[htbp]
  \centering
     \includegraphics[width=3.0in]{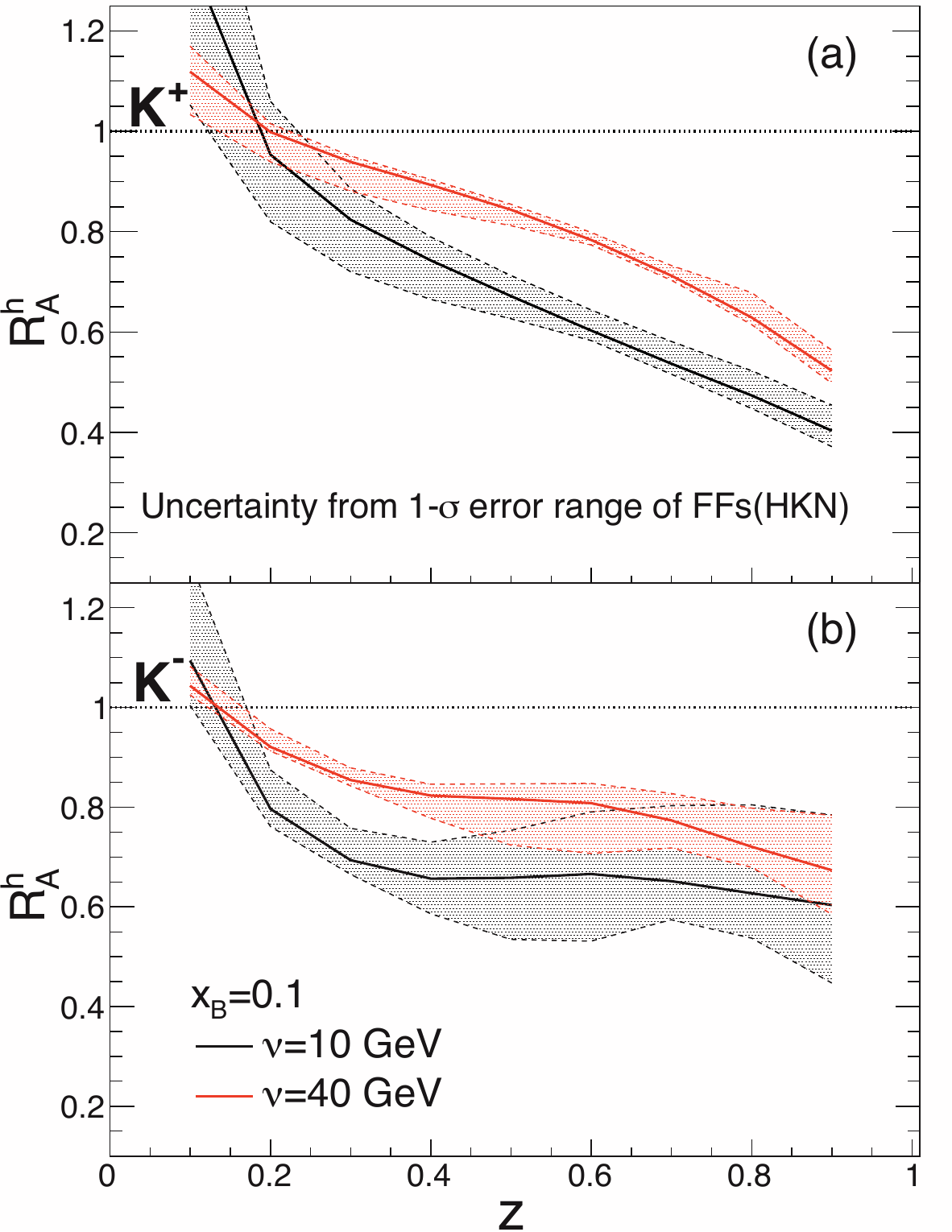}
   \caption{(color online) The nuclear modification factor for (a) $K^+$ and (b) $K^-$ for different initial quark energy $\nu$ in SIDIS off a Pb target at $x_B=0.1$ with HKN parameterization of vacuum FF's. Errors are from the HKN parameterization of vacuum FF's with 68\% CL propagated through the mDGLAP evolution equations.}
  \label{fig3}
\end{figure}
\end{center}

\begin{center}
\begin{figure}[htbp]
  \centering
     \includegraphics[width=3.0in]{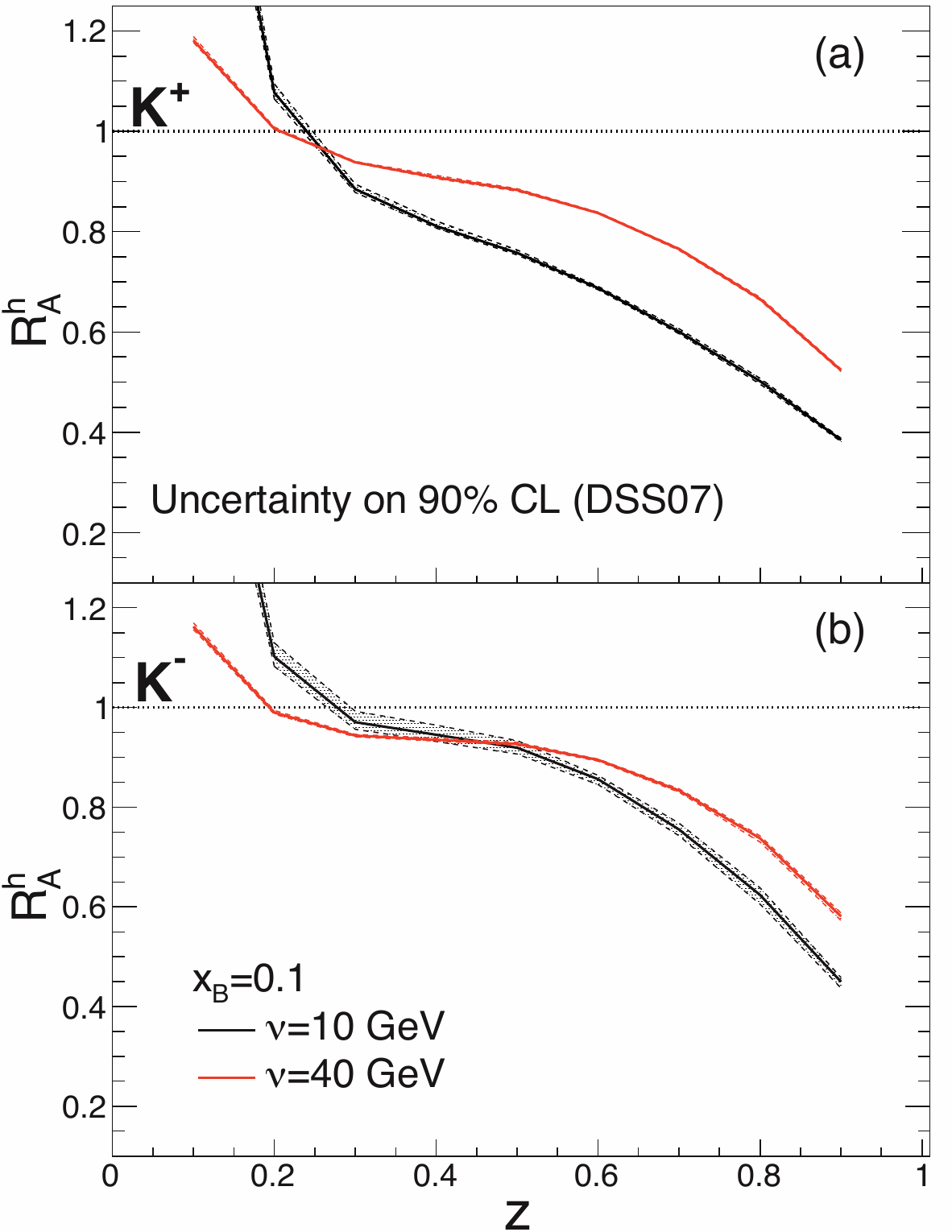}
   \caption{(color online) The nuclear modification factor for (a) $K^+$ and (b) $K^-$ for different initial quark energy $\nu$ in SIDIS off a Pb target at $x_B=0.1$
   with DSS07 parameterization of vacuum FF's. Errors are from the DSS07 parameterization of vacuum FF's with 90\% CL propagated through the mDGLAP evolution equations.}
  \label{fig4}
\end{figure}
\end{center}

\vspace{-0.5in}

In Figs.~\ref{fig3} and \ref{fig4}, we show the nuclear modification factor for charged kaons as a function of momentum fraction $z$ for different values of the initial quark energy $q^-=\nu$ at $x_B=0.1$ for HKN and DSS07 parameterizations of vacuum FF's, respectively. The kinematics considered in these two figures are similar to that in the HERMES experiment \cite{Airapetian:2007vu}. The spectra of both kaons are suppressed due to parton energy loss in the nuclear medium. The suppression decreases and eventually gives away to an enhancement at small $z$ due to the softening of the mFF's from radiated gluons and induced pair production. However, the suppression of $K^-$ is distinctly different from that of $K^+$ in part due to medium-induced flavor conversion. Quantitatively, the effect of medium-induced flavor conversion on the suppression factor for $K^-$ depends on the vacuum FF's that we use. With HKN parametrization, the $K^-$ suppression at large $z$ in Fig.~\ref{fig3} becomes flatter due to medium induced flavor conversion. For DSS07 parameterization, on the other hand, the suppression for $K^-$ in Fig.~\ref{fig4} is reduced only slightly at intermediate $z$ by the induced flavor conversion. This difference in the suppression factors for $K^+$ and $K^-$ is already present in the HERMES data \cite{Airapetian:2007vu,Chang:2014fba} and is a strong indication of the onset  of medium-induced flavor conversion.  The effect is, however, not very significant at $x_B\leq 0.1$ because there are still finite fractions of initial quarks with the constituent flavor of $K^-$ (see Table~\ref{tab1}) whose fragmentation dominates the $K^-$ spectrum at large $z$ in spite of their energy loss. Contributions from the induced flavor conversion in non-constituent initial quarks only offset partially the effect of energy loss of the initial constituent quarks and lead to a suppression factor for $K^-$ that is flatter than $K^+$.

In Figs.~\ref{fig3} and \ref{fig4}, we have included errors in the suppression factor for kaon spectra due to errors in the parameterization of vacuum FF's propagated through the mDGLAP evolution equations for mFF's.  In HKN parameterization \cite{HKN}, the Hessian method is used for the error analysis and errors for FF's from each parton flavor are considered independent of each other. Errors in the DSS07 \cite{deFlorian:2007aj,deFlorian:2014xna} parameterization are analyzed using the the method of Lagrange multipliers. With this method, errors for FF's of different parton flavors are correlated. This why errors from DSS07 parameterization on 90\% CL are much smaller than that from HKN parameterization on 68\% CL.

\begin{center}
\begin{figure}[htbp]
  \centering
     \includegraphics[width=3.0in]{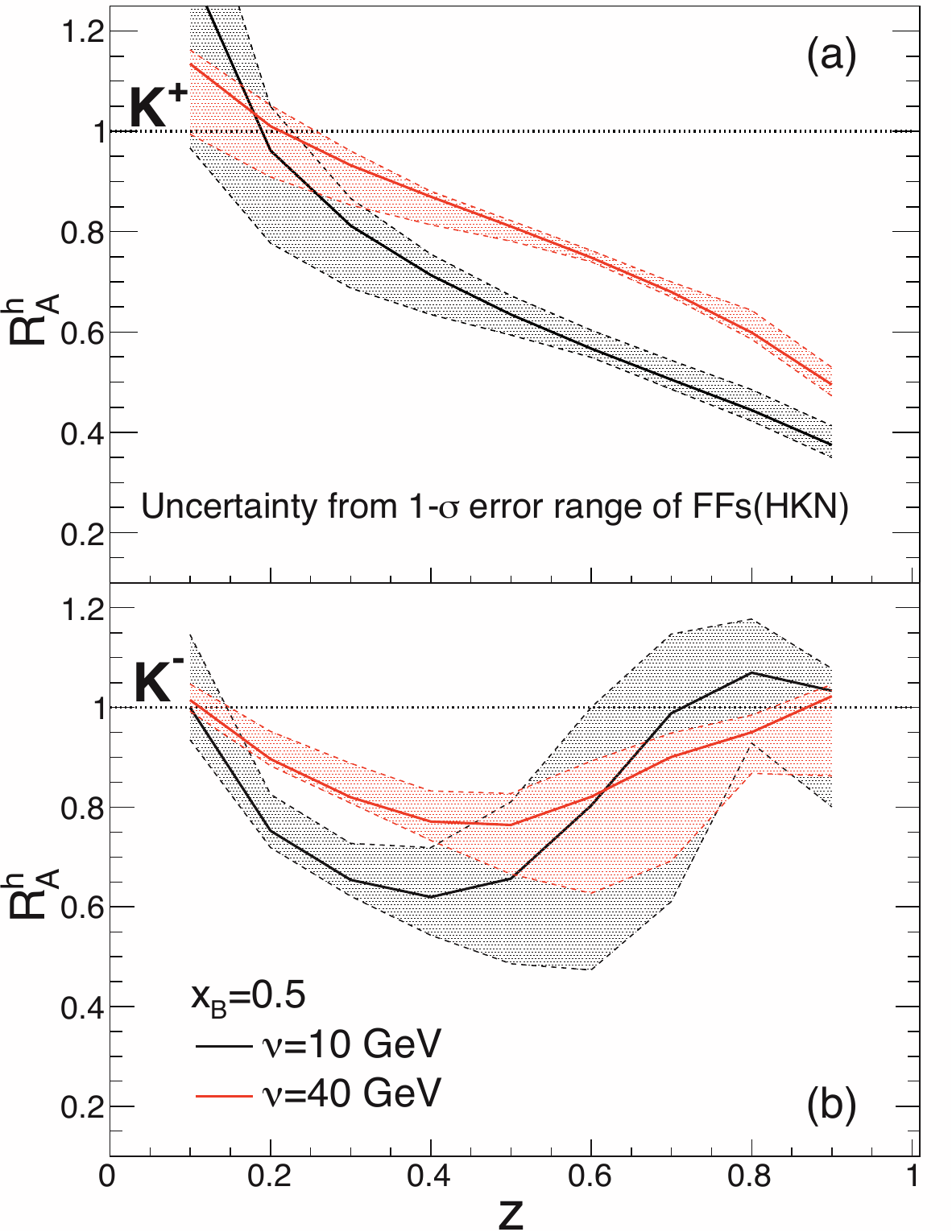}
  \caption{(color online) The same as Fig.~\ref{fig3} except for $x_B=0.5$.}
  \label{fig5}
\end{figure}
\end{center}

\begin{center}
\begin{figure}[htbp]
  \centering
     \includegraphics[width=3.0in]{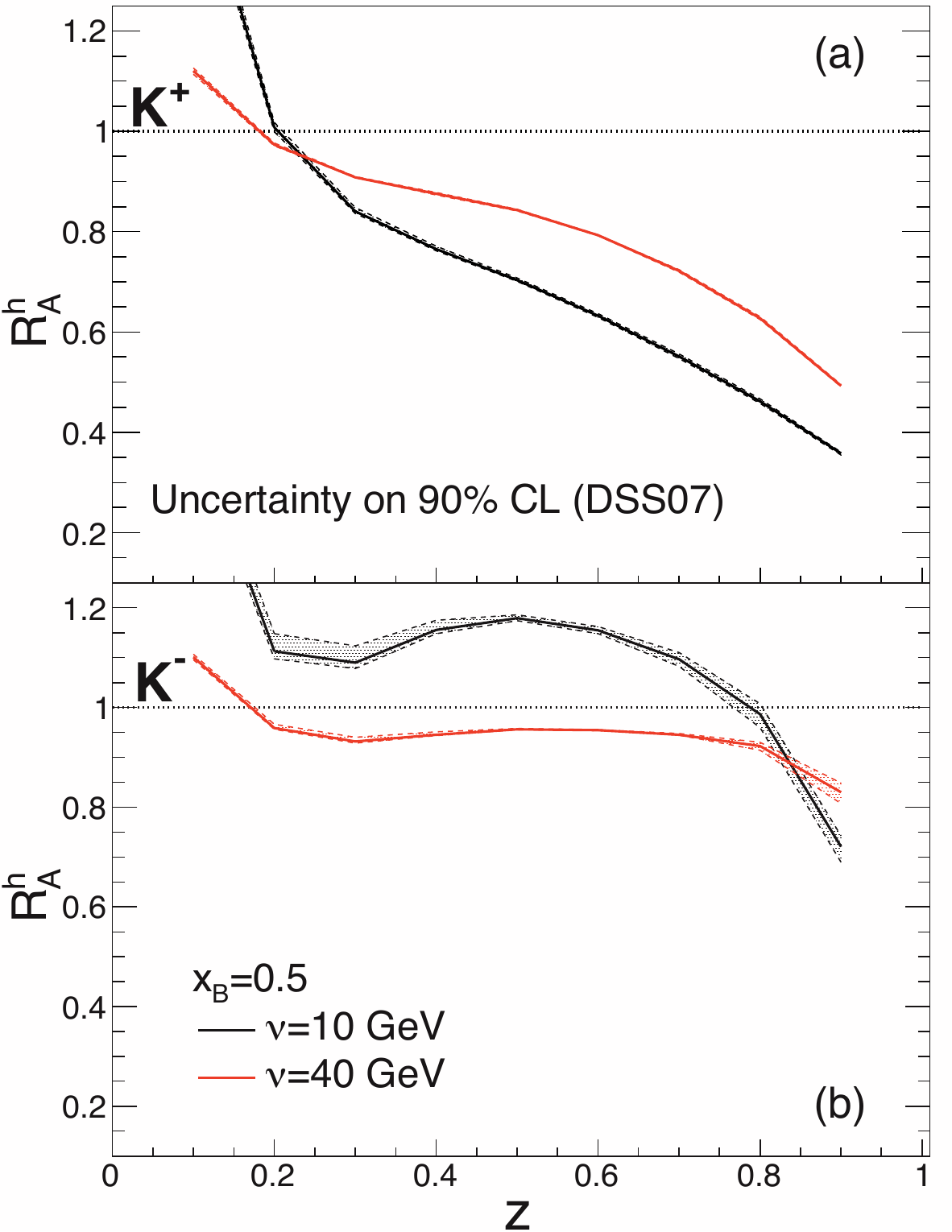}
  \caption{(color online) The same as  Fig.~\ref{fig4} except for $x_B=0.5$.}
  \label{fig6}
\end{figure}
\end{center}

In DIS events with larger values of $x_B$, the initial struck quarks are completely dominated by valence quarks of the nucleus. One should expect to see enhancement of $K^-$ due to medium-induced flavor conversion in the mFF's as shown in Fig.~\ref{fig2}. This is indeed confirmed in Figs.~\ref{fig5} and \ref{fig6} where we show nuclear modification factors for kaons at $x_B=0.5$ with HKN and DSS07 parameterizations of vacuum FF's, respectively. The suppression of $K^+$ is about the same as that at $x_B=0.1$ due to the weak dependence of medium modification on $Q^2$ \cite{Deng:2010xv,Chang:2014fba}.  The nuclear modification factor for $K^-$ is, however, completely  different and depends strongly on vacuum FF's. With HKN parameterization of vacuum FF's, the $K^-$ spectrum shown in Fig.~\ref{fig5} is suppressed at intermediate values of $z$ due to parton energy loss. But at large $z$, the modification factor starts to increase and approaches or exceeds 1, due to contributions from gluons and constituent quarks via medium-induced flavor conversion. With DSS07 parameterization of vacuum FF's, medium-induced flavor conversion also significantly reduces the effect of parton energy loss and can even lead to enhancement of $K^-$ at both intermediate and large values of $z$ as shown in Fig.~\ref{fig6}. The effect of medium-induced flavor conversion on $K^-$ spectra in both cases of vacuum FF's is stronger for smaller initial quark energy.

If we turn off the flavor conversion processes ($q\rightarrow gq$ and $g\rightarrow q\bar q$) in the mDGLAP equations, the rise of the nuclear modification factor for $K^-$ at large $z$ in Fig.~\ref{fig5} or intermediate and large $z$ in Fig.~\ref{fig6} disappears. Since both parton energy loss and flavor conversion are proportional to jet-medium interaction, the behavior of the modification factor for $K^-$ at intermediate and large $z$ is sensitive to the jet transport parameter $\hat q$. Experimental study of this phenomenon therefore provides another independent constraint on the jet transport parameter in the nuclear medium. The study can also be applied to $\bar K^0$ spectra. However, the effect in $K^0_S$ spectra will be reduced because it is a mixture of $K^0$ and $\bar K^0$.

In Figs.~\ref{fig3}-\ref{fig6}, we have included errors from each of the parameterizations of vacuum FF's which are propagated through mDGLAP equations for mFF's. One, however, should take these errors with a grain of salt. As illustrated in Ref.~\cite{Airapetian:2012ki}, none of these two parameterizations can fit perfectly the experimental data on $K^-$ spectra in SIDIS, though DSS07 can fit the $K^+$ spectra better.  These two different parameterizations of vacuum FF's for $K^-$ lead to quantitatively different nuclear modification factors for $K^-$. One should consider the difference as the largest theoretical uncertainty in our calculation of the nuclear modification factor which can only be reduced through better parameterization of vacuum FF's for $K^-$. 

\begin{center}
\begin{figure}[htbp]
  \centering
     \includegraphics[width=3.0in]{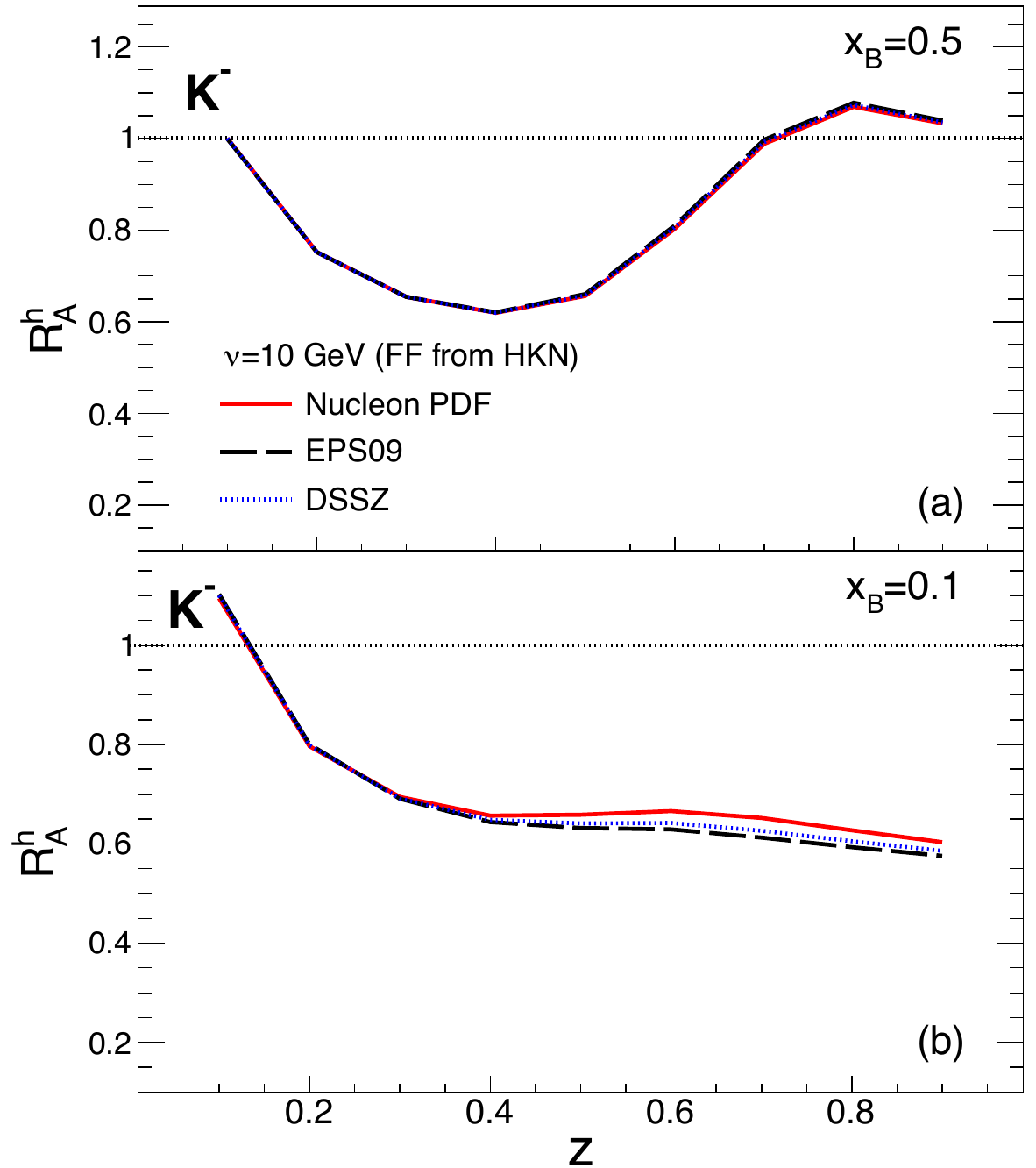}
  \caption{(color online) The nuclear modification factor for $K^-$ for an initial quark energy $\nu=10$ GeV in SIDIS off a Pb target at (a) $x_B=0.5$
and (b) $x_B=0.1$ with HKN parameterization of vacuum FF's and nucleon PDF's (solid), EPS09 (dashed) \cite{Eskola:2009uj} and DSSZ (dotted) \cite{deFlorian:2011fp} parameterization of nPDF's.}
  \label{fig7}
\end{figure}
\end{center}

\begin{center}
\begin{figure}[htbp]
  \centering
     \includegraphics[width=3.0in]{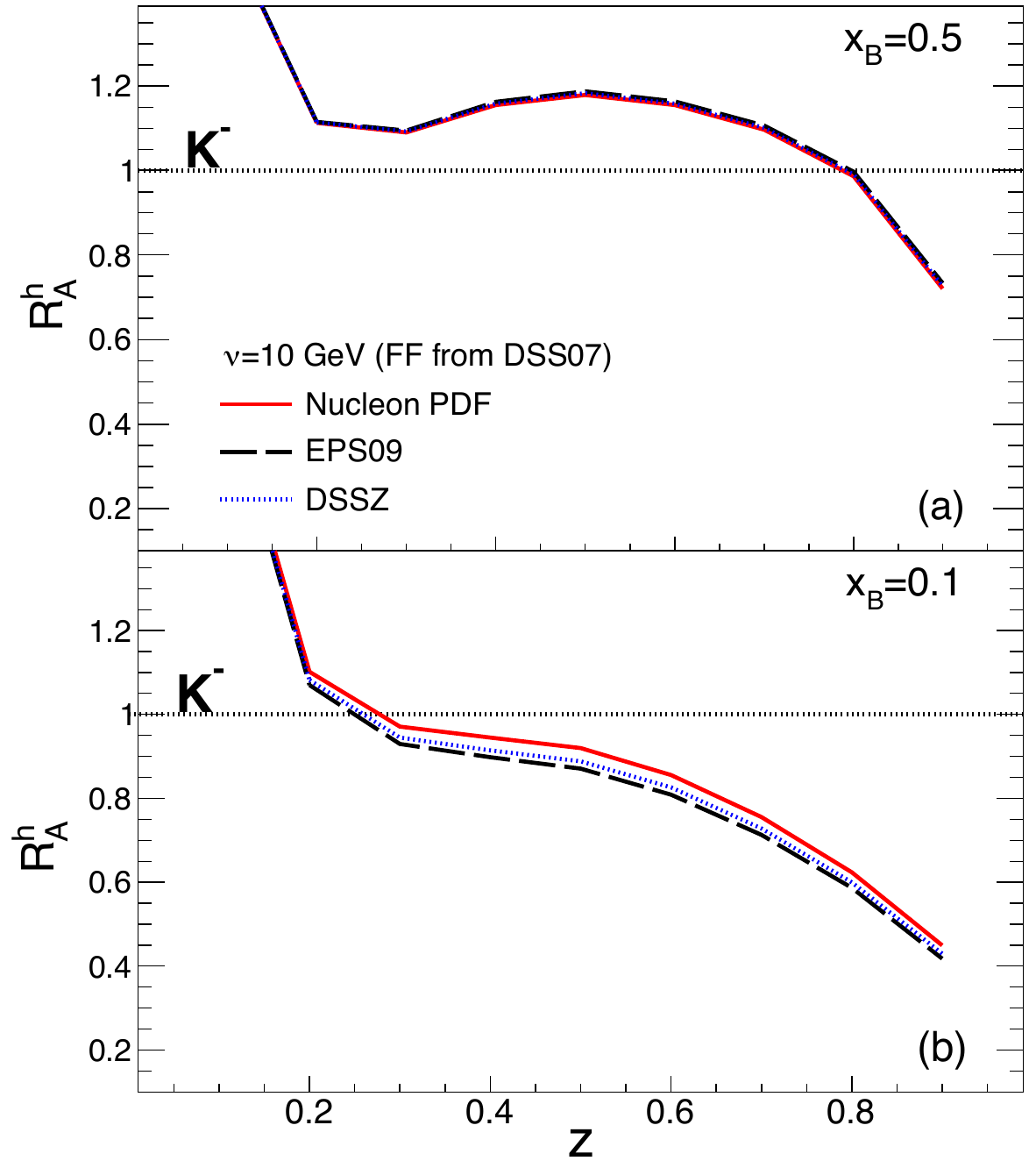}
  \caption{(color online) The same as  Fig.~\ref{fig7} except with DSS07 parameterization of vacuum FF's.}
  \label{fig8}
\end{figure}
\end{center}

Additional errors from the parton distribution functions should also be included in future analyses. One should also consider nuclear modification of parton distributions inside a nucleus. At intermediate and large $x_B$ region for our study here, the EMC effect \cite{Arneodo:1989sy} on nuclear modification of parton distributions should in principle depend on parton's flavor \cite{Cloet:2012td}. Therefore, it cannot be completely cancelled out in the hadron yield per DIS event as defined in Eq.~(\ref{yield-ratio}). Unfortunately, favor dependence of the EMC effect, especially for sea quarks and gluons, is not known. The difference in the EMC effects for sea quarks, valence quarks and gluons in existing parameterizations of nuclear parton distributions (nPDF) such as ESP09 \cite{Eskola:2009uj}  and DSSZ \cite{deFlorian:2011fp} are quite arbitrary and are only loosely constrained by momentum sum rule.  For the $K^-$ enhancement due to medium-induced flavor conversion in the large $x_B$ region we are interested in, the dominant contributions come from the favor conversion in the medium-modified fragmentation of struck valence quarks ($u$ and $d$). Since the flavor conversion in the mDGLAP evolution of the FFÕs is approximately independent of the favor of the initial valence quark, the EMC effect in nPDF will mostly cancel in the hadron yield per DIS event in Eq.~(\ref{yield-ratio}).   Shown in Figs.~\ref{fig7} and \ref{fig8}, are nuclear modification factors for $K^-$  calculated with HKN and DSS07 parameterizations of vacuum FF's, respectively, using nucleon PDF with proper isospin of the nucleus (solid),  ESP09 (dashed) and DSSZ (dotted) parameterizations of nPDFÕs.  The change to the nuclear modification factor for leading $K^-$ spectra in DIS at $x_B=0.5$ (upper panels)  due to the EMC effect is less than 1\%. At smaller $x_B=0.1$, contributions from sea quarks become sizable whose mFF's due to medium-induced flavor conversion and parton energy loss are somewhat different from that of valence quarks.  The nuclear effect in nPDF's  on the hadron yield per DIS event do not cancel completely. It reduces nuclear modification factor for $K^-$ by about 5\% as shown in Figs.~\ref{fig7}(b) and \ref{fig8}(b).

\section{Summary and discussion}
\label{sec:sum}

In conclusion, we have carried out a numerical study of the mDGLAP evolution of mFF's within a high-twist approach to parton propagation in medium. We discover that the nuclear modification factor for $K^-$ in SIDIS at large $x_B$ is very sensitive to the medium-induced flavor conversion. Instead of increased suppression like other hadrons (pions and $K^+$) due to parton energy loss of the leading quarks, the nuclear modification factor for $K^-$ is shown to rise and can exceed 1 at intermediate or/and large values of $z$ due to the proliferation of constituent quarks ($s$ and $\bar u$) and gluons from the induced flavor conversion to counter the effect of parton energy loss. This novel behavior is also found to be sensitive to the value of the jet transport parameter. Therefore, experimental measurements of such a phenomenon in a future electron-ion collider (EIC) can provide another independent probe of the properties of nuclear medium at high energies.  It can also help us to understand the flavor hierarchy of jet quenching phenomena in high-energy heavy-ion collisions \cite{Agakishiev:2011dc,OrtizVelasquez:2012te,Djordjevic:2013qba}.  According to our calculations presented in this article, effects of both parton energy loss on pions and $K^+$ and the medium-induced flavor conversion on $K^-$ spectra are most pronounced in SIDIS with small initial quark energy $\nu\sim 10$ GeV for a large nucleus target. This is also the lowest quark energy where the picture of quark propagation is still valid and one can approximately neglect effect of interaction between produced hadrons and nucleons inside the target.
A major numerical uncertainty in our quantitative  estimate of the effect of medium-induced flavor conversion on the final $K^-$ spectrum at intermediate or/and large $z$ is the parameterization of vacuum FF's for $K^-$.  Such uncertainty can also be reduced by future experimental measurements at EIC in the large $x_B$ regions.

{\bf Acknowledgments:} We thank M. Stratmann for communications and discussions about DSS07 parameterization of vacuum FF's and G.-Y. Qin for helpful discussions. This work is supported by the NSFC under Grant Nos.~11221504 and 11405066, China MOST under Grant No. 2014DFG02050, the Major State Basic Research Development Program in China (No. 2014CB845404),  by U.S. DOE under Contract No.~DE-AC02-05CH11231 and within the framework of the JET Collaboration.


\begin{thebibliography}{99}

\bibitem{Jacobs:2004qv} 
  P.~Jacobs and X.~N.~Wang,
  Prog.\ Part.\ Nucl.\ Phys.\  {\bf 54}, 443 (2005).
  
  
\bibitem{Majumder:2010qh} 
  A.~Majumder and M.~Van Leeuwen,
  Prog.\ Part.\ Nucl.\ Phys.\ A {\bf 66}, 41 (2011).
  
\bibitem{Bjorken:1982tu}
J.~D.~Bjorken, Fermilab Report No. FERMILAB-PUB-82/59-THY, 1982(unpublished)
  
\bibitem{Gyulassy:1993hr}
  M.~Gyulassy and X.~N.~Wang,
  Nucl.\ Phys.\  B {\bf 420}, 583 (1994).
  

\bibitem{Baier:1994bd}
  R.~Baier, Y.~L.~Dokshitzer, S.~Peigne and D.~Schiff,
  Phys.\ Lett.\  B {\bf 345}, 277 (1995);
  R.~Baier, Y.~L.~Dokshitzer, A.~H.~Mueller, S.~Peigne and D.~Schiff,
  Nucl.\ Phys.\  B {\bf 484}, 265 (1997).
  
\bibitem{Zakharov:1996fv}
  B.~G.~Zakharov,
  JETP Lett.\  {\bf 63}, 952 (1996).

\bibitem{Gyulassy:2000fs}
  M.~Gyulassy, P.~Levai and I.~Vitev,
  Phys.\ Rev.\ Lett.\  {\bf 85}, 5535 (2000);
  Nucl.\ Phys.\  B {\bf 594}, 371 (2001).
   
\bibitem{Wiedemann:2000za}
  U.~A.~Wiedemann,
  Nucl.\ Phys.\  B {\bf 588}, 303 (2000);
  Nucl.\ Phys.\  A {\bf 690}, 731 (2001).
  
\bibitem{Arnold:2001ba}
  P.~Arnold, G.~D.~Moore and L.~G.~Yaffe,
  JHEP {\bf 0111}, 057 (2001);
  JHEP {\bf 0206}, 030 (2002).
  
  \bibitem{Guo:2000nz}
  X.~F.~Guo and X.~N.~Wang,
  Phys.\ Rev.\ Lett.\  {\bf 85}, 3591 (2000);
  Nucl.\ Phys.\  A {\bf 696}, 788 (2001).

  
  
\bibitem{Wang:2002ri} 
  E.~Wang and X.~N.~Wang,
  Phys.\ Rev.\ Lett.\  {\bf 89}, 162301 (2002).
  
  
\bibitem{Arleo:2003jz} 
  F.~Arleo,
  Eur.\ Phys.\ J.\ C {\bf 30}, 213 (2003).

\bibitem{Ashman:1991cx} 
  J.~Ashman {\it et al.}  [European Muon Collaboration],
  Z.\ Phys.\ C {\bf 52}, 1 (1991).

\bibitem{Hafidi:2006ig} 
  K.~Hafidi [CLAS Collaboration],
  AIP Conf.\ Proc.\  {\bf 870}, 669 (2006).

  \bibitem{Airapetian:2007vu}
  A.~Airapetian {\it et al.}  [HERMES Collaboration],
  Nucl.\ Phys.\  B {\bf 780}, 1 (2007).

  
\bibitem{Burke:2013yra} 
  K.~M.~Burke
   {\it et al.} [JET Collaboration],
  Phys.\ Rev.\ C {\bf 90}, 014909 (2014).


\bibitem{Wang:2009qb}
  W.~T.~Deng and X.~N.~Wang,
  Phys.\ Rev.\  C {\bf 81}, 024902 (2010).

\bibitem{Deng:2010xv} 
  W.~T.~Deng, N.~B.~Chang and X.~N.~Wang,
  Nucl.\ Phys.\ A {\bf 855}, 416 (2011).
  
\bibitem{Chang:2014fba} 
  N.~B.~Chang, W.~T.~Deng and X.~N.~Wang,
  Phys.\ Rev.\ C {\bf 89}, 034911 (2014).
  
  
\bibitem{Liu:2006sf} 
  W.~Liu, C.~M.~Ko and B.~W.~Zhang,
  Phys.\ Rev.\ C {\bf 75}, 051901 (2007).


\bibitem{Liu:2008zb} 
  W.~Liu and R.~J.~Fries,
  Phys.\ Rev.\ C {\bf 77}, 054902 (2008).
  
  
\bibitem{Qiu:1990xy} 
  J.~W.~Qiu and G.~F.~Sterman,
  Nucl.\ Phys.\ B {\bf 353}, 137 (1991);
  \ B {\bf 353}, 105 (1991).
  
\bibitem{Kang:2013raa} 
  Z.~B.~Kang, E.~Wang, X.~N.~Wang and H.~Xing,
  Phys.\ Rev.\ Lett.\  {\bf 112}, 102001 (2014).
  
\bibitem{Kang:2014ela} 
  Z.~B.~Kang, E.~Wang, X.~N.~Wang and H.~Xing,
  arXiv:1409.1315 [hep-ph].
  
  
  \bibitem{Schafer:2007xh}
  A.~Schafer, X.~N.~Wang and B.~W.~Zhang,
  Nucl.\ Phys.\  A {\bf 793}, 128 (2007).
  

\bibitem{Majumder:2013re} 
  A.~Majumder,
  Phys.\ Rev.\ C {\bf 88}, 014909 (2013).
  
\bibitem{HKN}
  M.~Hirai, S.~Kumano, T.~H.~Nagai and K.~Sudoh,
  Phys.\ Rev.\  D {\bf 75}, 094009 (2007).

\bibitem{Airapetian:2012ki} 
  A.~Airapetian {\it et al.}  [HERMES Collaboration],
  Phys.\ Rev.\ D {\bf 87}, 074029 (2013).

\bibitem{deFlorian:2007aj} 
  D.~de Florian, R.~Sassot and M.~Stratmann,
  Phys.\ Rev.\ D {\bf 75}, 114010 (2007).
  
  
\bibitem{Pumplin:2002vw} 
  J.~Pumplin, D.~R.~Stump, J.~Huston, H.~L.~Lai, P.~M.~Nadolsky and W.~K.~Tung,
  JHEP {\bf 0207}, 012 (2002).
  
  
\bibitem{Airapetian:2013zaw} 
  A.~Airapetian {\it et al.}  [HERMES Collaboration],
  Phys.\ Rev.\ D {\bf 89}, no. 9, 097101 (2014).
  
\bibitem{deFlorian:2014xna} 
  D.~deFlorian, R.~Sassot, M.~Epele, R.~J.~Hern\'andez-Pinto and M.~Stratmann,
  Phys.\ Rev.\ D {\bf 91}, 014035 (2015).
  
\bibitem{Arneodo:1989sy} 
  M.~Arneodo {\it et al.}  [European Muon Collaboration],
  Nucl.\ Phys.\ B {\bf 333}, 1 (1990).


\bibitem{Cloet:2012td} 
  I.~C.~Cloet, W.~Bentz and A.~W.~Thomas,
  Phys.\ Rev.\ Lett.\  {\bf 109}, 182301 (2012).
  

\bibitem{Eskola:2009uj} 
  K.~J.~Eskola, H.~Paukkunen and C.~A.~Salgado,
  JHEP {\bf 0904}, 065 (2009).
  
\bibitem{deFlorian:2011fp} 
  D.~deFlorian, R.~Sassot, P.~Zurita and M.~Stratmann,
  Phys.\ Rev.\ D {\bf 85}, 074028 (2012).


\bibitem{Agakishiev:2011dc} 
  G.~Agakishiev {\it et al.}  [STAR Collaboration],
  Phys.\ Rev.\ Lett.\  {\bf 108}, 072302 (2012).

\bibitem{OrtizVelasquez:2012te} 
  A.~Ortiz Velasquez [ALICE Collaboration],
  Nucl.\ Phys.\ A {\bf 904-905}, 763c (2013).


\bibitem{Djordjevic:2013qba} 
  M.~Djordjevic and M.~Djordjevic,
  J.\ Phys.\ G {\bf 41}, 055104 (2014).


\end{thebibliography}
\end{document}